\documentclass[aps,prx,citeautoscript,twocolumn,showpacs,superscriptaddress]{revtex4-1}  

\usepackage{graphicx}  
\usepackage{dcolumn}   
\usepackage{bm}        
\usepackage{amssymb}   
\usepackage{amsmath}   
\usepackage{algorithm}
\usepackage{algpseudocode}
\usepackage[dvipsnames]{xcolor}
\usepackage{soul}
\usepackage{multirow}
\usepackage{tabularx}
\usepackage{array}
\usepackage{makecell}
\usepackage{chemformula}
\usepackage{refstyle}

\usepackage[normalem]{ulem}

\begin{document}

\title{DQC: a Python program package for Differentiable Quantum Chemistry}

\author{M. F. Kasim}
\email{Present address: Machine Discovery Ltd, muhammad@machine-discovery.com}
\affiliation{Department of Physics, Clarendon Laboratory, University of Oxford, Parks Road, Oxford OX1 3PU, UK}
\author{S. Lehtola}
\affiliation{Molecular Sciences Software Institute, Blacksburg, Virginia 24061, United States}
\author{S. M. Vinko}
\email{sam.vinko@physics.ox.ac.uk}
\affiliation{Department of Physics, Clarendon Laboratory, University of Oxford, Parks Road, Oxford OX1 3PU, UK}
\affiliation{Central Laser Facility, STFC Rutherford Appleton Laboratory, Didcot OX11 0QX, UK}

\date{\today}

\begin{abstract}

Automatic differentiation represents a paradigm shift in scientific programming, where evaluating both functions and their derivatives is required for most applications. By removing the need to explicitly derive expressions for gradients, development times can be be shortened, and calculations simplified.
For these reasons, automatic differentiation has fueled the rapid growth of a variety of sophisticated machine learning techniques over the past decade, but is now also increasingly showing its value to support {\it ab initio} simulations of quantum systems, and enhance computational quantum chemistry.
Here we present an open-source differentiable quantum chemistry simulation code, DQC, and explore applications facilitated by automatic differentiation: (1) calculating molecular perturbation properties; (2) reoptimizing a basis set for hydrocarbons; (3) checking the stability of self-consistent field wave functions; and (4) predicting molecular properties via alchemical perturbations.

\end{abstract}


\maketitle

\section{Introduction}

Automatic differentiation is a collection of techniques used to evaluate, up to machine precision, the derivative of a function specified by a computer program. It allows software developers to focus solely on designing the best model for a given problem, without having to worry about implementing any derivatives of the model with respect to its various mathematical parameters.
It has already had a transformative effect in machine learning, enabling the development of many new techniques over the past decade, such as batch normalization~\cite{ioffe2015batchnorm}, attention layers~\cite{vaswani2017attention}, and unique neural network architectures~\cite{schutt2018schnet,ronneberger2015unet}.

Automatic differentiation is still relatively new in the context of computational sciences, but is already showing promise across a diverse set of applications including tensor networks~\cite{liao2019differentiable-tensor-network}, computational fluid dynamics~\cite{schenck2018spnets-diff-cfd}, and molecular dynamics simulations~\cite{schoenholz2020jaxmd}.
Automatic differentiation is also growing increasingly popular in quantum chemistry, where it has been used to optimize molecular basis sets~\cite{tamayo2018automatic-diffiqult}, to calculate higher derivatives of various exchange-correlation (xc) functionals~\cite{ekstrom2010arbitrary-xc} of density functional theory (DFT)~\cite{hohenberg1964inhomogeneous, kohn1965self-ks}, and to determine arbitrary-order nuclear coordinate derivatives of electronic energies~\cite{abbott2021arbitrary-geom-qc}.

Automatic differentiation is also an essential stepping stone to enable direct integration of quantum chemistry methods with machine learning models and their training.
In this context, a differentiable implementation of DFT was recently used to learn the xc functional~\cite{li2021kohn-sham-regularizer} from accurate reference calculations within the density matrix renormalization group approach, or from a mixture of computational and experimental data~\cite{kasim2021learning-xc}, showing a promising new approach to developing transferable and robust xc functionals via deep learning.

Although differentiation in quantum chemistry can be done via finite-difference schemes, calculating the gradient of a parameter with respect to some other input parameter can be time consuming as one has to run the simulation as many times as the number of input parameters.
Moreover, finite-difference methods are prone to numerical instabilities and are very sensitive to the chosen step-size. Automatic differentiation addresses these challenges by calculating the analytical gradient via the chain rule, eliminating both the need to run the simulation many times, and the need for step-size tuning.

To address the growing need for automatic differentiation in quantum chemistry, we introduce Differentiable Quantum Chemistry (DQC), a DFT and Hartree--Fock (HF)~\cite{hartree1928wave} simulation code.
DQC is written in Python using PyTorch~\cite{neurips2019-pytorch} and xitorch~\cite{kasim2020xitorch}.
While PyTorch provides gradient calculations for elementary operations such as matrix multiplication and explicit eigendecomposition, xitorch provides gradient calculations for functional operations such as root finding, optimization, and implicit eigendecomposition.
The use of PyTorch and xitorch in DQC enables various applications in quantum chemistry that would otherwise be far more difficult to pursue. For example, the work by Kasim and Vinko~\cite{kasim2021learning-xc} on learning the xc functional directly from a set of heterogeneous experimental data and calculated density profiles within the constraints of Kohn--Sham DFT already mentioned above was enabled by the use of DQC.

We begin this paper by outlining the basic theory behind DQC in section~\ref{sec:methods}. The implementation is described in section~\ref{sec:implementations}.
Applications that exemplify the present approach are given in section~\ref{sec:apps}.
We discuss the challenges in the use of automatic differentiation techniques in computational science in section~\ref{sec:discussions}, and summarize our work in section~\ref{sec:conclusions}.


\section{Methods}
\label{sec:methods}

Quantum chemical calculations of the electronic structure typically require the evaluation of abstract functionals, such as root finding for self-consistent field (SCF) iterations, and minimization for geometry optimizations and the direct minimization approach to the SCF problem. We discuss the handling of these functionals in the context of DQC in this section.

\subsection{SCF iterations}

DQC is based on the use of a Gaussian basis set within the linear combination of atomic orbitals approach (LCAO).
For simplicity, we will only present the theory for the spin-restricted formalism, as the spin-unrestricted (and restricted open-shell) formalisms are analogous.
The SCF iterations proceed in the LCAO approach by solving Roothaan's equation~\cite{roothaan1951new, lehtola2020overview}
\begin{equation}
    \label{eq:roothaan}
    \mathbf{F(D)C} = \mathbf{SCE},
\end{equation}
where $\mathbf{F}(\mathbf{D})$ is the Fock matrix which is a function of the density matrix $\mathbf{D}$, $\mathbf{C}$ is the orbital matrix, $\mathbf{S}$ is the overlap matrix, and $\mathbf{E}$ is a diagonal matrix containing the orbital energies.
The generalized eigenvalue equation in \eqref{roothaan} can be reduced to the normal form by orthogonalizing the overlap matrix $\mathbf{S}$~\cite{lowdin1956quantum}, and operating in the linearly independent basis spanned by the eigenvectors of $\mathbf{S}$ with large enough eigenvalues; any ill-conditioning in the overlap matrix can be removed by choosing a well-defined sub basis with the help of a pivoted Cholesky decomposition~\cite{lehtola2019curing}.
The density matrix $\mathbf{D}$, required to construct the Fock matrix $\mathbf{F}$, can be obtained by
\begin{equation}
    \label{eq:dm}
    \mathbf{D} = \mathbf{CNC^\dagger}
\end{equation}
with $\mathbf{N}$ a diagonal matrix containing the occupation numbers of the orbitals.
As discussed in ref.~\citenum{lehtola2020overview}, the SCF procedure requires repeatedly solving \eqref{roothaan, dm} until self-consistency is achieved.

The Roothaan iteration in \eqref{roothaan} can be written mathematically as the process of finding a vector $\mathbf{y}$ such that
\begin{equation}
    \label{eq:self-consistent}
    \mathbf{y} = \mathbf{f}(\mathbf{y}, \boldsymbol{\theta}),
\end{equation}
where $\mathbf{f}$ is a function that takes the vector $\mathbf{y}$ and other parameters $\boldsymbol{\theta}$, and returns the expected value of vector $\mathbf{y}$.
In DQC, the parameter $\mathbf{y}$ is represented by the Fock matrix $\mathbf{F}$, so the function $\mathbf{f}$ is the procedure that solves equation (\ref{eq:roothaan}), calculating the density matrix from equation (\ref{eq:dm}) and constructing back the Fock matrix from the density matrix.
The parameters $\boldsymbol{\theta}$ represent the other variables involved in the calculation, such as the overlap matrix $\mathbf{S}$ and the occupation number matrix $\mathbf{N}$.
The algorithm and gradient calculation for \eqref{roothaan, self-consistent} are already available in PyTorch and xitorch.

\subsection{Direct minimization}

An alternative approach to solving the self-consistent field equations is to directly find the orbitals that minimize the total energy $\mathcal{E}$ by the use of optimization algorithms~\cite{head1988optimization}.
The energy $\mathcal{E}$ can be calculated from the density matrix $\mathbf{D}$ which can be obtained in turn from the orbital coefficients $\mathbf{C}$ using \eqref{dm}.
This relation makes the energy a function of the orbital matrix, $\mathcal{E}(\mathbf{C})$.
However, as the orbitals must remain orthonormal, $\mathbf{C^\dagger SC} = \mathbf{I}$, we introduce a new variable $\mathbf{Q}$, defined in terms of its relation with $\mathbf{C}$ as
\begin{align}
    \label{eq:uq-relation-1}
    \mathbf{Q} &= \mathbf{\hat{Q}R} \\
    \label{eq:uq-relation-2}
    \mathbf{C} &= \mathbf{S}^{-1/2} \mathbf{\hat{Q}}.
\end{align}
\Eqref{uq-relation-1} is the QR decomposition of the matrix $\mathbf{Q}$ into an orthogonal matrix $\mathbf{\hat{Q}}$ and an upper triangular matrix $\mathbf{R}$.
\Eqref{uq-relation-2} involves the inverse square root of the overlap matrix $\mathbf{S}$, which can be computed using the eigendecomposition of $\mathbf{S}$.
The energy can then be parameterized by $\mathbf{Q}$, reducing the direct minimization problem to an unbounded optimization problem:
\begin{equation}
    \label{eq:variational-q-problem}
    \mathbf{Q^*} = \arg\min_\mathbf{Q} \mathcal{E}(\mathbf{Q}).
\end{equation}
The gradient of the energy with respect to $\mathbf{Q}$, i.e. $\partial \mathcal{E}/\partial \mathbf{Q}$, is required for an efficient solution. It is automatically computed by PyTorch and xitorch.

Once the optimum $\mathbf{Q}^*$ in equation (\ref{eq:variational-q-problem}) is found, $\mathbf{Q}^*$ can still be differentiated with respect to any other variables, such as the nuclear coordinates or the occupation number matrix $\mathbf{N}$.
This is made possible by the gradient of the optimization functional provided by xitorch.

We note that our approach of using QR decomposition is slightly different from common direct minimization techniques that use the matrix exponential of a skew-Hermitian matrix such as the geometric direct minimization algorithm of ref.~\citenum{van2002geometric-variational}.

\newcommand{\dsetsmallspacing}{-2.0ex}
\newcommand{\dsetsmallspacingafter}{-2.0ex}
\newcommand{\dsetcolwidth}{0.34\linewidth}
\begin{table}[t]
    \caption{Execution speed comparison between DQC (SCF iterations) and PySCF.}
    \label{tab:speed}
    \centering
    \begin{ruledtabular}
    \begin{tabular}{lrr}
        Cases &
        DQC &
        PySCF \\ [\dsetsmallspacingafter] \\
        
        \hline \\ [\dsetsmallspacing]
        H$_2$O (HF/cc-pVDZ) & $96\ \mathrm{ms}$ & $245\ \mathrm{ms}$ \\
        H$_2$O (PW92/cc-pVDZ) & $530\ \mathrm{ms}$ & $430\ \mathrm{ms}$ \\
        C$_4$H$_5$N (HF/cc-pVTZ) & $108\ \mathrm{s}$ & $17\ \mathrm{s}$ \\
        C$_4$H$_5$N (PW92/cc-pVTZ) & $101\ \mathrm{s}$ & $25\ \mathrm{s}$ \\
        C$_4$H$_5$N (density fit PW92/cc-pVTZ) & $30\ \mathrm{s}$ & $22\ \mathrm{s}$ \\
        C$_6$H$_8$O$_6$ (density fit PW92/cc-pVDZ) & $87\ \mathrm{s}$ & $57\ \mathrm{s}$ \\
        [\dsetsmallspacingafter] \\
    \end{tabular}
    \end{ruledtabular}
\end{table}

\section{Implementation}
\label{sec:implementations}

DQC implements restricted and unrestricted HF and Kohn--Sham DFT calculations without periodic boundary conditions.
The energy can be minimized using either SCF iterations with Broyden's good method~\cite{broyden1965class} for Fock matrix updates, or with direct minimization using gradient descent with momentum~\cite{pearlmutter1991gd-momentum}; more elaborate SCF convergence accelerators will be implemented at a later stage of the project.
All parameters are differentiable with respect to any other parameter present in the calculation, including nuclear coordinates, atomic numbers, electron occupation numbers, as well as the basis set exponents and contraction coefficients. The differentiability of these parameters allows for the exploration of new applications with DQC, a few of which are presented in what follows.

Although execution speed is not a top priority for DQC, the program shows good overall performance.
\Tabref{speed} compares the running times of DQC with PySCF~\cite{sun2020recent-pyscf}, an established quantum chemistry code.
For small systems, we find that DQC is as efficient as PySCF.
However, for moderate-sized molecules (e.g. C$_4$H$_5$N), DQC is about 4--6 times slower than PySCF.
This slowdown can be attributed to DQC currently not taking advantage of the symmetry of the 2-electron integral tensor, which can reduce the tensor size by roughly a factor of 8 when the basis functions are real-valued.
In contrast, DQC is only about 50\% slower than PySCF for systems with density fitting. The difference is mainly caused by the higher number of SCF iterations required in DQC over PySCF, due to the use of different SCF convergence acceleration schemes in the two codes.

\begin{figure}
    \centering
    \includegraphics[width=\linewidth]{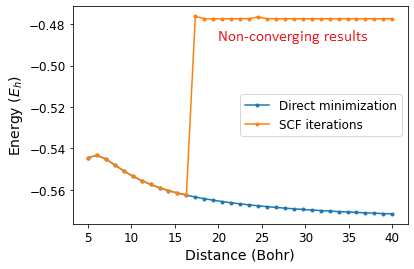}
    \caption{Unrestricted PW92/cc-pVDZ~\cite{perdew1992accurate-pw92, dunning1989gaussian-ccbasis} energy for the hydrogen molecule cation H$_2^+$ as a function of the internuclear distance calculated via the SCF (using Broyden's method~\cite{broyden1965class}) and direct minimization approaches. Note that the SCF iterations do not converge for internuclear distances larger than 17 Bohr.}
    \label{fig:variational-vs-scf}
\end{figure}

\section{Applications}
\label{sec:apps}

\subsection{Direct minimization}
\label{subsec:direct-minimization-app}

The automatic availability of gradients makes it easy to implement the direct minimization method in DQC.
Direct minimization is known to be more robust than SCF iterations in finding a converged solution~\cite{lehtola2020overview}, and is particularly useful in difficult cases such as for molecules near their dissociation limits.
We illustrate this in \figref{variational-vs-scf}, where we show the PW92/cc-pVDZ~\cite{perdew1992accurate-pw92, dunning1989gaussian-ccbasis} total energy of H$_2^+$ as a function of the internuclear distance.
The SCF method does not converge for internuclear distances greater than around 17 Bohr. In contrast, the direct minimization method continues to converge even for substantially larger distances in excess of 40 Bohr.

\subsection{Checking SCF stability}
\label{subsec:stability-app}

One difficulty in SCF calculations is that the calculation can converge onto a saddle point, which corresponds to an excited state~\cite{lehtola2020overview}.
If the calculation has converged onto a local minimum, the Hessian of the energy with respect to orbital rotations must be positive semidefinite. Saddle point solutions, instead, correspond to one or more negative eigenvalues of the orbital Hessian.
To maintain the orthonormality of the orbitals, the Hessian is calculated based on the $\mathbf{Q}$ variable.
Conveniently, we do not need to derive the expression for the Hessian matrix $\partial^2\mathcal{E}/\partial \mathbf{Q}^2$ as it is automatically obtained by PyTorch.

Moreover, only the lowest eigenvalue needs to be calculated for the stability check, so the full Hessian matrix does not need to be constructed. We obtain the lowest eigenvalue using Davidson's iterative algorithm~\cite{davidson1975iterative}, as implemented in xitorch.
Note that the gradients of the lowest energy eigenvalue with respect to all other parameters in the calculation are automatically available, which may prove useful in further future applications.

\begin{table}[t]
    \caption{Lowest eigenvalue of the HF/pc-1 orbital Hessian matrix. All of the calculations correspond to stationary points of the HF energy.}
    \label{tab:scf-stability}
    \centering
    \begin{ruledtabular}
    \begin{tabular}{lrrr}
        Molecules &
        Distance&
        Energy &
        Lowest eigenvalue \\
        & (Bohr) & ($E_h$) & ($E_h$)\\
        [\dsetsmallspacingafter] \\
        
        \hline \\ [\dsetsmallspacing]
        O$_2$ (ground) & 2.0 & $-149.54$ & $-6\times 10^{-14}$ \\
        O$_2$ (excited) & 2.0 & $-149.17$ & $-0.73$ \\
        BeH (ground) & 2.5 & $-15.14$ & $-4\times 10^{-14}$ \\
        BeH (excited) & 2.5 & $-15.01$ & $-0.25$ \\
        CH (ground) & 2.0 & $-38.259$ & $-5\times 10^{-8}$ \\
        CH (excited) & 2.0 & $-38.256$ & $-0.07$ \\
        [\dsetsmallspacingafter] \\
    \end{tabular}
    \end{ruledtabular}
\end{table}

We show some examples of SCF stability checks for HF/pc-1~\cite{jensen2001} calculations on diatomic molecules in \tabref{scf-stability}.
As can be seen from the data, the lowest eigenvalues of the orbital Hessian are very close to zero for the ground state, while the excited states yield large negative eigenvalues. This illustrates that it is straightforward to check whether or not a state corresponds to a true minimum with DQC.

\subsection{Molecular properties}
\label{subsec:molprop-app}

A key advantage of writing quantum chemistry software with automatic differentiation is that calculations of molecular properties can be implemented efficiently:
all we need to know is how a specific property relates to some derivative expression.
For example, vibrational modes and frequencies of a molecule can be written as
\begin{equation}\label{eq:vibr}
    \mathbf{q}_{\rm vib}, \omega_{\rm vib} = \mathrm{eig}\left(\frac{\partial^2 \mathcal{E}}{\partial \mathbf{X}^2}\right),
\end{equation}
where $\mathbf{X}$ is an $n\times 3$ matrix containing the nuclear coordinates of the $n$ atoms, $\mathrm{eig}$ is the eigendecomposition, $\mathbf{q}_{\rm vib}$ is one of the vibrational modes, and $\omega_{\rm vib}$ is its frequency.
The expression shown in \eqref{vibr} is all that is needed to calculate the vibrational characteristics of the molecule; the explicit form of the derivative expression is not required.

\begin{table}[t]
    \caption{Perturbative properties of H$_2$O from HF/cc-pVDZ calculations. The middle column presents the values calculated in DQC while the last column shows the CCCBDB values~\cite{NIST_CCCBDB}. The IR intensities and Raman activities are presented at the frequency of 1800 cm$^{-1}$.}
    \label{tab:perturbative-prop}
    \centering
    \begin{ruledtabular}
    \begin{tabular}{lrr}
        Properties &
        DQC &
        CCCBDB \\ [\dsetsmallspacingafter] \\
        
        \hline \\ [\dsetsmallspacing]
        IR intensities (km/mol) & 80.69 & 80.70 \\
        Raman activities (A$^4$/amu) & 4.79 & 4.79 \\
        Dipole (D) & $-2.044$ & $-2.044$ \\
        Quadrupole$_{xx}$ (DA) & $-7.008$ & $-7.008$ \\
        [\dsetsmallspacingafter] \\
    \end{tabular}
    \end{ruledtabular}
\end{table}

Another useful example is the calculation of the electric dipole and quadrupole moments of a molecule. The electric dipole moment is given by
\begin{equation} \label{eq-dipole}
    \boldsymbol{\mu} = -\frac{\partial \mathcal{E}}{\partial \mathbf{E}} + \sum_i Z_i \mathbf{x}_i,
\end{equation}
while the electric quadrupole tensor is given by
\begin{equation}
    \mathbf{M} = -2\frac{\partial \mathcal{E}}{\partial \nabla\mathbf{E}} + \sum_i Z_i \mathbf{x}_i\mathbf{x}_i^T.
\end{equation}
In both expressions, $\mathcal{E}$ is the total energy of the molecule, $\mathbf{E}$ is the electric field, $\mathbf{Z}_i$ is the atomic number of $i$-th atom, and $\mathbf{x}_i$ are its coordinates.

Having these vibrational and electric multipole properties readily available makes obtaining the Raman vibrational spectrum straightforward.
For example, the intensity of the infrared vibrational transition for a normal mode $\mathbf{q}$ at a frequency $\omega$ is given by
\begin{equation}
    I_{\rm IR} = \sum_i\left(\sum_{jk} \frac{\partial \mu_i}{\partial X_{jk}} q_{jk}\right)^2,
\end{equation}
where $\mathbf{\mu}$ is the electric dipole moment and $\mathbf{X}$ is the matrix of atomic coordinates.

Similarly, the Raman activity at the same frequency and normal mode is proportional to~\cite{p2007analytic-raman}
\begin{align}
    I_{\rm Raman} &= 5 \left[\mathrm{tr}\left(\boldsymbol{\alpha}\right)\right]^2 + 7 \gamma; \\
    \alpha_{ij} &= \sum_{kl}\frac{\partial^2 \mu_i}{\partial E_j\partial X_{kl}} q_{kl}; \\
    \gamma &= \sum_{ij}(1 - \delta_{ij})\left[\frac{1}{4}(\alpha_{ii} - \alpha_{jj})^2 + \frac{3}{2}\alpha_{ij}^2\right], \label{eq-activity}
\end{align}
where $\delta_{ij}$ is the Kronecker delta.

\begin{figure}
    \centering
    \includegraphics[width=\linewidth]{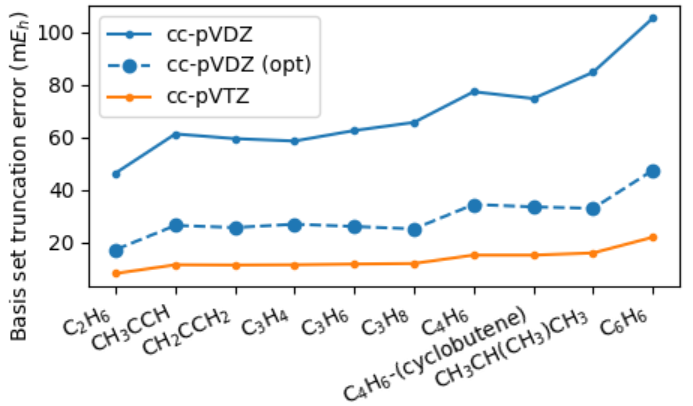}
    \caption{Basis set truncation errors for the cc-pVDZ and cc-pVTZ basis sets for a range of hydrocarbons, compared with the truncation error for a reoptimized cc-pVDZ basis set. The reference energies are computed in the cc-pV5Z basis set. None of the molecules shown in the figure were used in the basis set optimization.}
    \label{fig:basis-opt}
\end{figure}

The HF/cc-pVDZ perturbation properties of H$_2$O, found using the expressions above, are displayed in Table \ref{tab:perturbative-prop}. 
The values are in excellent agreement with data from the the computational chemistry comparison and benchmark database (CCCBDB)~\cite{NIST_CCCBDB}, even though DQC does not have any explicit code to calculate the gradients required for these properties.

\subsection{Basis set optimization}
\label{subsec:basopt-app}

The differentiability of DQC with respect to the basis set parameters enables the optimization of system-specific basis sets.
Here we show this capability by optimizing a basis set for hydrocarbons within Kohn--Sham DFT using the PW92 functional.
We reoptimized the cc-pVDZ basis~\cite{dunning1989gaussian-ccbasis} for a training set of molecules consisting of CH (methylidyne), CH$_3$ (methyl radical), CH$_4$ (methane), C$_2$H$_2$ (acetylene), and C$_2$H$_4$ (ethylene).
The accuracy of the reoptimized basis was then tested on a set of hydrocarbons that were not included in the training set. The results are shown in \figref{basis-opt}. The reoptimization of the cc-pVDZ basis set leads to a marked decrease of the total energies of all the molecules in the test set, as the cc-pVXZ basis set series is designed to capture correlation energies instead of polarization energies~\cite{dunning1989gaussian-ccbasis}, and as hydrocarbons are chemically similar.

\subsection{Alchemical perturbation}
\label{subsec:alchemy-app}

One of the differentiable quantities in DQC is the atomic number, allowing us to perform alchemical perturbation studies to predict the properties of molecules without actually needing to simulate them~\cite{balawender2019exploring-alchemical, von2020alchemical-dft}.
As a simple example, we will show here that some properties of diatomic molecules CO (atomic numbers 6 and 8) and BF (atomic numbers 5 and 9) can be estimated directly from the properties of N$_2$ (atomic number 7) and its alchemical perturbations.
To do this, we parametrize the atomic number of the atoms in the diatomic molecule by a continuous variable $\lambda$, so that the atoms have atomic numbers $Z_l = 7 + \lambda$ and $Z_r = 7 - \lambda$.
The molecules N$_2$, CO, and BF thus correspond to $\lambda=0$, $\lambda=1$, and $\lambda=2$, respectively.

The properties we aim to predict are the bond length $s^*$ and the energy $E^*$ at the equilibrium position, which can be mathematically expressed as
\begin{align}
    s^*(\lambda) &= \arg\min_s E(s, \lambda), \\
    E^*(\lambda) &= E(s^*, \lambda).
\end{align}
Performing HF/pc-1 calculations, we evaluate the equilibrium distance and the energy at the equilibrium position in two ways.
The first way is to optimize the geometry for various fixed values of $\lambda$, and calculate the above properties directly.
The calculations were performed separately, but with the same basis (pc-1 nitrogen basis on all atoms) for different molecules, in analogy to the procedure used in refs~\citenum{balawender2019exploring-alchemical} and \citenum{von2020alchemical-dft}.
The second way is to estimate those properties using a Taylor series expansion to second order in $\lambda$:
\begin{align}
    s^*(\lambda) &\approx s^*(0) + \lambda \frac{\partial s^*}{\partial \lambda}(0) + \frac{1}{2}\lambda^2 \frac{\partial^2 s^*}{\partial \lambda^2}(0) \\
    E^*(\lambda) &\approx E^*(0) + \lambda \frac{\partial E^*}{\partial \lambda}(0) + \frac{1}{2}\lambda^2 \frac{\partial^2 E^*}{\partial \lambda^2}(0).
\end{align}
As $s^*$ and $E^*$ are calculated at equilibrium, calculating the perturbation terms requires propagating the gradient through the optimization process. However, automatic differentiation makes the propagation trivial, as it is automatically handled by the optimization routine in xitorch.

\begin{figure}
    \centering
    \includegraphics[width=\linewidth]{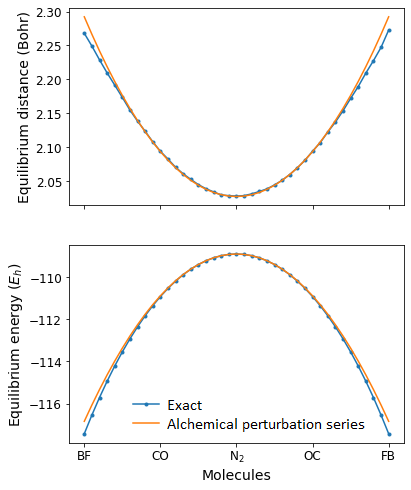}
    \caption{Comparison of the properties of CO and BF obtained via an exact calculation and an estimation from the second-order gradients via alchemical perturbations, employing the nitrogen pc-1 basis on all atoms.}
    \label{fig:alchemy}
\end{figure}

The results obtained via these two approaches are compared in \figref{alchemy}.
As we can see from these results, the properties of CO and BF can be estimated accurately from alchemical perturbations of N$_2$.
The estimated equilibrium distances for CO and BF differ by $-0.0004$ and 0.024 Bohr, respectively, while the estimated equilibrium energies deviate by 33 and 587 m$E_h$, respectively.
This shows that the equilibrium position of new molecules can be estimated well with the alchemical gradient calculated by DQC, without actually having to calculate those molecules.

\section{Discussion}
\label{sec:discussions}

Implementing quantum chemistry with automatic differentiation libraries is a promising way to accelerate simulation workflows and to enable novel applications. However, the implementation comes with several challenges.
An overarching challenge is that the automatic differentiation library used here, PyTorch, is primarily designed for deep learning rather than for scientific computing. As deep learning only focuses on low-order derivatives, accessing high-order gradients that are commonly required for scientific applications can be difficult.

Detecting numerical instabilities in high-order gradient calculations can also be demanding.
Instabilities that produce NaN (not-a-number) in PyTorch are relatively straightforward to manage with its debugging feature since version 1.8, but other instabilities that do not produce a NaN can be challenging to detect.

Another challenge is debugging and profiling higher-order gradient calculations.
As the gradient is generated automatically, it is hard to find the slowest part of the code, or the part that requires the most memory, because this information is not readily provided by the available profiling tools.

Besides higher-order gradient calculations, using automatic differentiation libraries also poses unique challenges in terms of code optimization.
For example, quantum chemistry codes usually save the two-electron integrals on disk due to their large size and process them only in blocks small enough to fit easily into memory.
This scheme can only be used in DQC if the gradients with respect to the nuclear positions and the basis are not required.
To the authors' best knowledge, there is currently no obvious structure in PyTorch to allow gradient-propagating operations to work with large tensors saved on disk.

\section{Conclusions}
\label{sec:conclusions}

Implementing quantum chemical calculations using automatic differentiation liberates us from needing to derive analytical gradient expressions.
With gradients automatically generated by the program, software developers can focus on designing better, more detailed computational models, and on applying them to problems at hand.  
We have shown how automatic differentiation allows us to easily explore various applications in quantum chemistry, and are confident that further exploration of this approach will unveil new applications that have not been considered to date.

\section*{Code availability}

The Differentiable Quantum Chemistry (DQC) code can be found at https://github.com/diffqc/dqc/.
The repository that contains the applications presented in this paper is located at https://github.com/diffqc/dqc-apps/.

\begin{acknowledgments}
M.F.K. and S.M.V. acknowledge support from the UK EPSRC grant EP/P015794/1 and the Royal Society.
S.M.V. is a Royal Society University Research Fellow.
The authors declare no conflict of interest.
\end{acknowledgments}

\bibliographystyle{unsrt}
\bibliography{bibliography}

\begin{thebibliography}{10}

\bibitem{ioffe2015batchnorm}
Sergey Ioffe and Christian Szegedy.
\newblock Batch normalization: Accelerating deep network training by reducing
  internal covariate shift.
\newblock In {\em International conference on machine learning}, pages
  448--456. PMLR, 2015.

\bibitem{vaswani2017attention}
Ashish Vaswani, Noam Shazeer, Niki Parmar, Jakob Uszkoreit, Llion Jones,
  Aidan~N Gomez, Lukasz Kaiser, and Illia Polosukhin.
\newblock Attention is all you need.
\newblock {\em arXiv preprint arXiv:1706.03762}, 2017.

\bibitem{schutt2018schnet}
Kristof~T Sch{\"u}tt, Huziel~E Sauceda, P-J Kindermans, Alexandre Tkatchenko,
  and K-R M{\"u}ller.
\newblock Schnet--a deep learning architecture for molecules and materials.
\newblock {\em J. Chem. Phys.}, 148(24):241722, 2018.

\bibitem{ronneberger2015unet}
Olaf Ronneberger, Philipp Fischer, and Thomas Brox.
\newblock U-net: Convolutional networks for biomedical image segmentation.
\newblock In {\em International Conference on Medical image computing and
  computer-assisted intervention}, pages 234--241. Springer, 2015.

\bibitem{liao2019differentiable-tensor-network}
Hai-Jun Liao, Jin-Guo Liu, Lei Wang, and Tao Xiang.
\newblock Differentiable programming tensor networks.
\newblock {\em Physical Review X}, 9(3):031041, 2019.

\bibitem{schenck2018spnets-diff-cfd}
Connor Schenck and Dieter Fox.
\newblock Spnets: Differentiable fluid dynamics for deep neural networks.
\newblock In {\em Conference on Robot Learning}, pages 317--335. PMLR, 2018.

\bibitem{schoenholz2020jaxmd}
Samuel Schoenholz and Ekin~Dogus Cubuk.
\newblock Jax md: a framework for differentiable physics.
\newblock {\em Advances in Neural Information Processing Systems}, 33, 2020.

\bibitem{tamayo2018automatic-diffiqult}
Teresa Tamayo-Mendoza, Christoph Kreisbeck, Roland Lindh, and Al{\'a}n
  Aspuru-Guzik.
\newblock Automatic differentiation in quantum chemistry with applications to
  fully variational hartree--fock.
\newblock {\em ACS central science}, 4(5):559--566, 2018.

\bibitem{ekstrom2010arbitrary-xc}
Ulf Ekstr\"{o}m, Lucas Visscher, Radovan Bast, Andreas~J Thorvaldsen, and
  Kenneth Ruud.
\newblock Arbitrary-order density functional response theory from automatic
  differentiation.
\newblock {\em Journal of chemical theory and computation}, 6(7):1971--1980,
  2010.

\bibitem{hohenberg1964inhomogeneous}
Pierre Hohenberg and Walter Kohn.
\newblock Inhomogeneous electron gas.
\newblock {\em Phys. Rev.}, 136(3B):B864, 1964.

\bibitem{kohn1965self-ks}
Walter Kohn and Lu~Jeu Sham.
\newblock Self-consistent equations including exchange and correlation effects.
\newblock {\em Phys. Rev.}, 140(4A):A1133, 1965.

\bibitem{abbott2021arbitrary-geom-qc}
Adam~S Abbott, Boyi~Z Abbott, Justin~M Turney, and Henry~F Schaefer~III.
\newblock Arbitrary-order derivatives of quantum chemical methods via automatic
  differentiation.
\newblock {\em The Journal of Physical Chemistry Letters}, 12(12):3232--3239,
  2021.

\bibitem{li2021kohn-sham-regularizer}
Li~Li, Stephan Hoyer, Ryan Pederson, Ruoxi Sun, Ekin~D Cubuk, Patrick Riley,
  and Kieron Burke.
\newblock {Kohn-Sham equations as regularizer: Building prior knowledge into
  machine-learned physics}.
\newblock {\em Phys. Rev. Lett.}, 126(3):036401, 2021.

\bibitem{kasim2021learning-xc}
Muhammad~F Kasim and Sam~M Vinko.
\newblock Learning the exchange-correlation functional from nature with fully
  differentiable density functional theory.
\newblock {\em arXiv preprint arXiv:2102.04229}, 2021.

\bibitem{hartree1928wave}
Douglas~Rayne Hartree.
\newblock The wave mechanics of an atom with a non-coulomb central field. part
  ii. some results and discussion.
\newblock In {\em Mathematical Proceedings of the Cambridge Philosophical
  Society}, volume~24, pages 111--132. Cambridge University Press, 1928.

\bibitem{neurips2019-pytorch}
Adam Paszke, Sam Gross, Francisco Massa, Adam Lerer, James Bradbury, Gregory
  Chanan, Trevor Killeen, Zeming Lin, Natalia Gimelshein, Luca Antiga, Alban
  Desmaison, Andreas Kopf, Edward Yang, Zachary DeVito, Martin Raison, Alykhan
  Tejani, Sasank Chilamkurthy, Benoit Steiner, Lu~Fang, Junjie Bai, and Soumith
  Chintala.
\newblock Pytorch: An imperative style, high-performance deep learning library.
\newblock In H.~Wallach, H.~Larochelle, A.~Beygelzimer, F.~d\textquotesingle
  Alch\'{e}-Buc, E.~Fox, and R.~Garnett, editors, {\em Advances in Neural
  Information Processing Systems}, volume~32, pages 8026--8037. Curran
  Associates, Inc., 2019.

\bibitem{kasim2020xitorch}
Muhammad~F Kasim and Sam~M Vinko.
\newblock $\xi$-torch: differentiable scientific computing library.
\newblock {\em arXiv preprint arXiv:2010.01921}, 2020.

\bibitem{roothaan1951new}
Clemens Carel~Johannes Roothaan.
\newblock New developments in molecular orbital theory.
\newblock {\em Rev. Mod. Phys.}, 23(2):69, 1951.

\bibitem{lehtola2020overview}
Susi Lehtola, Frank Blockhuys, and Christian Van~Alsenoy.
\newblock An overview of self-consistent field calculations within finite basis
  sets.
\newblock {\em Molecules}, 25(5):1218, 2020.

\bibitem{lowdin1956quantum}
Per-Olov L{\"o}wdin.
\newblock Quantum theory of cohesive properties of solids.
\newblock {\em Adv. Phys.}, 5(17):1--171, 1956.

\bibitem{lehtola2019curing}
Susi Lehtola.
\newblock {Curing basis set overcompleteness with pivoted Cholesky
  decompositions}.
\newblock {\em J. Chem. Phys.}, 151(24):241102, 2019.

\bibitem{head1988optimization}
Martin Head-Gordon and John~A Pople.
\newblock {Optimization of wave function and geometry in the finite basis
  Hartree--Fock method}.
\newblock {\em J. Phys. Chem.}, 92(11):3063--3069, 1988.

\bibitem{van2002geometric-variational}
Troy Van~Voorhis and Martin Head-Gordon.
\newblock A geometric approach to direct minimization.
\newblock {\em Molecular Physics}, 100(11):1713--1721, 2002.

\bibitem{broyden1965class}
Charles~G Broyden.
\newblock A class of methods for solving nonlinear simultaneous equations.
\newblock {\em Mathematics of computation}, 19(92):577--593, 1965.

\bibitem{pearlmutter1991gd-momentum}
Barak~A Pearlmutter.
\newblock Gradient descent: Second-order momentum and saturating error.
\newblock {\em Advances in neural information processing systems}, pages
  887--894, 1991.

\bibitem{sun2020recent-pyscf}
Qiming Sun, Xing Zhang, Samragni Banerjee, Peng Bao, Marc Barbry, Nick~S Blunt,
  Nikolay~A Bogdanov, George~H Booth, Jia Chen, Zhi-Hao Cui, et~al.
\newblock {Recent developments in the PySCF program package}.
\newblock {\em J. Chem. Phys}, 153(2):024109, 2020.

\bibitem{perdew1992accurate-pw92}
John~P Perdew and Yue Wang.
\newblock Accurate and simple analytic representation of the electron-gas
  correlation energy.
\newblock {\em Phys. Rev. B}, 45(23):13244, 1992.

\bibitem{dunning1989gaussian-ccbasis}
Thom~H Dunning~Jr.
\newblock {Gaussian basis sets for use in correlated molecular calculations. I.
  The atoms boron through neon and hydrogen}.
\newblock {\em J. Chem. Phys.}, 90(2):1007--1023, 1989.

\bibitem{davidson1975iterative}
E.~R. Davidson.
\newblock The iterative calculation of a few of the lowest eigenvalues and
  corresponding eigenvectors of large real symmetric matrices.
\newblock {\em Journal of Computational Physics}, 17(1):87--94, 1975.

\bibitem{jensen2001}
Frank Jensen.
\newblock {Polarization consistent basis sets: Principles}.
\newblock {\em J. Chem. Phys.}, 115:9113--9125, 2001.

\bibitem{NIST_CCCBDB}
{Editor: R. D. Johnson III}.
\newblock {NIST Computational Chemistry Comparison and Benchmark Database. NIST
  Standard Reference Database Number 101 (Release 21, August 2020), [Online].
  Available: {\tt{https://dx.doi.org/10.18434/T47C7Z}}. National Institute of
  Standards and Technology, Gaithersburg, MD.}, 2020.

\bibitem{p2007analytic-raman}
Darragh P.~O'Neill, Mih{\'a}ly K{\'a}llay, and J{\"u}rgen Gauss.
\newblock Analytic evaluation of raman intensities in coupled-cluster theory.
\newblock {\em Molecular Physics}, 105(19-22):2447--2453, 2007.

\bibitem{balawender2019exploring-alchemical}
Robert Balawender, Michael Lesiuk, Frank De~Proft, Christian Van~Alsenoy, and
  Paul Geerlings.
\newblock Exploring chemical space with alchemical derivatives: alchemical
  transformations of h through ar and their ions as a proof of concept.
\newblock {\em Physical Chemistry Chemical Physics}, 21(43):23865--23879, 2019.

\bibitem{von2020alchemical-dft}
Guido~Falk von Rudorff and O~Anatole von Lilienfeld.
\newblock Alchemical perturbation density functional theory.
\newblock {\em Physical Review Research}, 2(2):023220, 2020.

\end{thebibliography}

\end{document}